\documentclass[%
 reprint,
nofootinbib,
 amsmath,amssymb,
 aps,
 pra,
floatfix,
]{revtex4-2}

\usepackage{graphicx}
\usepackage{dcolumn}
\usepackage{bm}
\usepackage{cancel}
\usepackage{xcolor}
\usepackage{hyperref}
\hypersetup{colorlinks=true,allcolors = blue}
\usepackage{cleveref}

\begin{document}

\preprint{APS/123-QED} 

\title{
Quantum states resembling classical periodic trajectories in mesoscopic elliptic billiards
} 


\author{Jes\'{u}s G. Riestra}
\author{Julio C. Guti\'{e}rrez-Vega}
\email{juliocesar@tec.mx}
\affiliation{Photonics and Mathematical Optics Group, Tecnol\'{o}gico de Monterrey, Monterrey, M\'{e}xico 64849}

\date{\today}

\begin{abstract}

A quantum wave function with localization on classical periodic orbits in a mesoscopic elliptic billiard has been achieved by appropriately superposing nearly degenerate eigenstates expressed as products of Mathieu functions. We analyze and discuss the rotational and librational regimes of motion in the elliptic billiard. Simplified line equations corresponding to the classical trajectories can be extracted from the quantum state as an integral equation involving angular Mathieu functions. The phase factors appearing in the integrals are connected to the classical initial positions and velocity components. We analyze the probability current density, the phase maps, and the vortex distributions of the periodic orbit quantum states for both rotational and librational motions; furthermore, they may represent traveling and standing trajectories inside the elliptic billiard.  
\end{abstract}

\maketitle


\noindent J. G. Riestra and J. C. Guti\'errez-Vega, \textit{Quantum states resembling classical periodic trajectories in mesoscopic elliptic billiards}, Phys. Rev. E \textbf{109}, 034205 (2024).
\href{https://doi.org/10.1103/PhysRevE.109.034205}{https://doi.org/10.1103/PhysRevE.109.034205}

\section{Introduction} 

The problem of two-dimensional (2D) billiards consists of a point-like
particle moving inside a planar closed domain, bouncing elastically at its
boundary \cite{TabachnikovBOOK}. The dynamics of the particle can be studied in both classical and quantum regimes \cite{GutzwillerBOOK}.
Despite the apparent simplicity of the billiard system, it provides a
means of exploring a wide range of physical phenomena that can be extrapolated
to more complex systems. For instance, periodic stable trajectories, energy spectra, chaoticity, optical-quantum analogies, quantum dots, quantum confinement effects, and quantum-classical connections, among other phenomena, can be investigated by studying billiards \cite{TabachnikovBOOK,GutzwillerBOOK,KozlovBOOK}. The shape of the boundary highly determines the billiard behavior. Symmetric
boundaries, such as rectangular, circular, or elliptical, tend to be
integrable systems with two constants of motion, promoting separability \cite{Berry1981,Waalkens,pollett1995elliptic,VanZon}. 
Conversely, more irregular boundaries usually lead to non-integrable
systems with varying levels of chaos, e.g., Sinai or stadium billiards \cite{GutzwillerBOOK,KozlovBOOK}.

Under some favorable conditions, integrable billiards allow finding analytically the characteristic equations to obtain classical periodic trajectories. To establish a quantum-classical connection, these periodic orbits can be related to quantum wave functions in analogy with ballistic transport in quantum systems, ray light distributions in waveguides, intracavity fields in optical resonators, etc. 
Actually, the connection between classical trajectories and quantum coherent states has been studied recurrently over the years since the introduction of the coherent states of the harmonic oscillator by E. Schr{\"o}dinger \cite{schrodinger1926stetige}. The suitable superposition of degenerate states of the 2D harmonic oscillator produces coherent states with minimum uncertainty that resemble the expected classical trajectories of a particle moving under the action of the two-dimensional isotropic parabolic potential 
\cite{wodkiewicz1985coherent,buvzek1989generalized,de1992oscillator,Chen2005}. 
The SU(2) representation has also been used to construct stationary coherent states localized on Lissajous figures in the 2D quantum harmonic oscillator with commensurate frequencies \cite{Chen2003HO,makowski2005comment,Chen2005}.

Within the context of billiards, the classical-quantum connection between classical periodic orbits and superpositions of quantum states was recently studied for free particles confined in square, equilateral triangular, and circular billiards
\cite{Chen2002,chen2003vortex,Liu2006,hsieh2017extracting}. 
In these cases, the quantum state consists of the superposition of nearly degenerate eigenstates where the degeneracy condition is connected with the classical parameters.
Due to the relatively simple analytical solutions of these highly symmetric billiards, explicit equations of the classical trajectories could be extracted from the quantum eigenstates \cite{chen2023systematically}.

It is known that the elliptic billiard is an integrable non-trivial system with two constants of motion \cite{pollett1995elliptic, Zhang, Waalkens, P05, P16, P123}, namely, the energy $E$ and the product of angular momenta about the foci $\Gamma$. The particle has two regimes of motion depending on the sign of the second constant of motion: rotational ($\mathcal{R}$-type) for positive $\Gamma,$ and librational ($\mathcal{L}$-type) for negative $\Gamma$. Elliptic billiards have proven beneficial in creating models for elliptical micro-cavity lasers \cite{spiridonov}. These elliptical structures are fabricated to break the rotational symmetry of circular resonators and, thus, manipulate the directionality of the output radiation \cite{spiridonov2017}. A more circular resonator produces whispering gallery modes (R-type), while a more elliptic cavity emits radiation mostly along its axes (L-type). Therefore, it is essential to characterize the possible wave distributions within elliptic cavities for applications.

In this paper, we determine the quantum states that closely resemble classical periodic orbits of a free particle in an elliptic billiard. Throughout the paper, these states will be referred to as \textit{Periodic-Orbit Quantum States} (POQS). They are constructed with suitable superpositions of nearly degenerate eigenstates of the elliptic billiard with specific amplitude and phase coefficients. The conditions for near degeneracy are described in detail and involve a very accurate computation of the radial Mathieu functions, their characteristic values, and parametric zeros. Although the even and odd eigenstates of the quantum elliptic billiard are not degenerate, especially in the case of the librational motion, we will show that an efficient wave localization on the classical trajectories can be achieved.

Using the analytical expansion of the eigenstates in terms of plane waves, we could extract the classical trajectory equations from the POQS in the form of integral line equations. Our study reveals that the phase factors involved in the superposition are related to the classical position and velocity. 
Additionally, we characterize the POQS's quantum probability current. In the case of $\mathcal{R}$-type trajectories, we found that the probability current aligns with the classical velocities, and a series of unit-charge vortices emerge along the interfocal line. On the other hand, for the $\mathcal{L}$-type trajectories, the vector current has some discrepancies with the classical velocities, which we will discuss in detail. 

This work extends and consolidates previous studies of the connections between quantum states and classical periodic trajectories in 2D integrable billiards \cite{Chen2002,chen2003vortex,Liu2006,hsieh2017extracting,chen2023systematically}. Our analysis contributes to the global understanding of the properties of elliptical billiards and establishes a link between its classical and quantum description that had not been studied before, as far as we know. 

\section{Classical periodic orbits and quantum eigenstates of the elliptic billiard}

We will briefly describe the classical trajectories and the quantum
eigenstates of a particle in the elliptic billiard to establish notation and
provide necessary formulas \cite{Waalkens, P05, P16, P123}. 

Consider a point particle of mass $M$ moving inside an elliptic boundary given by 
\begin{equation}
    x^{2}/a^{2}+y^{2}/b^{2}=1, \qquad b\leq a,
\end{equation}
and whose foci are located at $x_{\pm}=\pm f=\pm\left(a^{2}-b^{2}\right)^{1/2}$. The
eccentricity of the ellipse is $\epsilon=f/a\in\lbrack0,1).$

The dynamics of the particle is conveniently described in elliptic coordinates
$\mathbf{r}=(\xi,\eta)$ defined by
\begin{equation}
x=f\cosh\xi\cos\eta,\qquad y=f\sinh\xi\sin\eta,
\end{equation}
where $\xi\in\lbrack0,\xi_{0}]$ and $\eta\in(-\pi,\pi]$ are the
\textit{radial} and \textit{angular} elliptic coordinates, respectively. The
lower limit $\xi=0$ corresponds to the interfocal line $\left\vert x\right\vert
\leq f,$ and the upper limit $\xi=\xi_{0}=\mathrm{arctanh}\left(  b/a\right)
$ defines the boundary of the billiard. The scaling factors of the elliptic
coordinates are%
\begin{equation}
h_{\bot} \equiv h_{\xi}=h_{\eta} = f\sqrt{(\cosh 2\xi - \cos 2\eta)/2}.
\end{equation}

As it moves in the billiard, the particle has two constants of motion \cite{Zhang}. The first one is the energy
\begin{equation}
E=\frac{\mathbf{p}\cdot\mathbf{p}}{2M}=\frac{p_{\xi}^{2}+p_{\eta}^{2}}%
{2Mh_{\bot}^{2}},
\end{equation}
and the second one is the dot product of the angular momenta about the foci of
the ellipse
\begin{equation}
\Gamma=\mathbf{L}_{1}\cdot\mathbf{L}_{2}=\frac{f^{2}}{h_{\bot}^{2}}(p_{\eta}^{2}%
\sinh^{2}\xi-p_{\xi}^{2}\sin^{2}\eta), \label{LambdaEquation}%
\end{equation}
where $p_{\xi}$ and $p_{\eta}$ are the canonical momenta in elliptic
coordinates. Because there exist two constants of motion, the particle is
restricted to move in a specific trajectory in the phase space, where $E$ and $\Gamma$ do not change.

For a given energy $E,$ the parameter $\Gamma$ lies within the interval
$\Gamma\in\lbrack-2MEf^{2},2MEb^{2}].$ Thus, it is convenient to define the non-dimensional constant of motion \cite{P16}
\begin{equation}
\gamma\equiv\frac{\Gamma}{2MEf^{2}}\in\left[  -1,\frac{b^{2}}{f^{2}}\right]  .
\label{gama}%
\end{equation}
whose range depends only on the geometric parameters of the billiard.

\subsection{Classical periodic trajectories}

\begin{figure}[t]
\includegraphics[width=8cm]{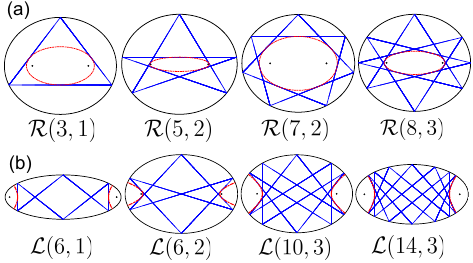}
\caption{(a) $\mathcal{R}$-type periodic classic orbits for several indices
$(p,n)$ in an elliptic billiard with eccentricity $\epsilon=0.5.$ Each
trajectory is tangent to an elliptic caustic displayed with a dash-dotted red line. (b)
$\mathcal{L}$-type trajectories with $\gamma_{p,n}<0$. All segments of the
trajectory cross the $x$-axis through the interfocal line. Eccentricities
are $\epsilon=0.924,$ $\epsilon=0.714,$ and $\epsilon=0.848$. Dash-dotted red lines are the hyperbolic caustics of the librational orbits. }%
\label{F1_fig:tray_clas}%
\end{figure}

As illustrated in Fig. \ref{F1_fig:tray_clas}, the particle in the elliptic billiard presents two kinds of motion:
\begin{itemize}
\item Rotational ($\mathcal{R}$-type) when $\gamma>0$. The particle rotates
around the interfocal line, crossing the $x$ axis outside the foci. All
segments of the trajectory are tangent to a confocal elliptic caustic given by
$\xi_{C}=\mathrm{arccosh}\left(  \sqrt{1+\gamma}\right)  ,$ as shown in Fig. \ref{F1_fig:tray_clas}(a). The radial coordinate $\xi$ is restricted to the range $\xi\in\lbrack
\xi_{C},\xi_{0}]$, and the angular one is unrestricted.

\item Librational ($\mathcal{L}$-type) when $\gamma<0.$ The particle bounces alternately between the top and bottom of the ellipse, crossing the $x$ axis through the interfocal line; see Fig. \ref{F1_fig:tray_clas}(b). The particle is confined between
two hyperbolic caustics defined by $\eta=\pm\eta_{C}$ and $\eta = \pm\left(
\pi-\eta_{C}\right)  $, where $\eta_{C}=\mathrm{arccos}\left(  \sqrt{1+\gamma
}\right)  \in(0,\pi/2).$
\end{itemize}

The value $\gamma=0$ corresponds to the separatrix between rotational and
librational motions. In this case, the path segments alternately pass through
the foci of the ellipse and tend to align with the $x$-axis as they
successively bounce off the boundary. The minimum value $\gamma=-1$
corresponds to the vertical motion along the $y$-axis bouncing alternatively
at the covertex points of the ellipse at $y=\pm b$. The maximum value
$\gamma=b^{2}/f^{2}$ corresponds to the limiting rotational path that runs along the elliptic boundary.

\begin{figure*}[t]
\includegraphics[width=17cm]{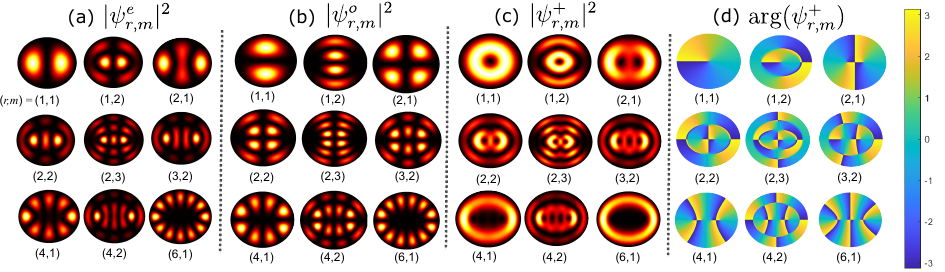}
\caption{Lower-order eigenstates in the elliptic billiard with eccentricity
$\epsilon=0.5.$ (a) $|\psi_{r,m}^{e}(\xi,\eta)|^{2}$ of even standing-wave
states. (b) $|\psi_{r,m}^{o}(\xi,\eta)|^{2}$ of odd standing-wave states. (c)
$|\psi_{r,m}^{+}(\xi,\eta)|^{2}$ of positive traveling-wave states. (d) Phase
distribution $\arg\psi_{r,m}^{+}(\xi,\eta)$ of positive traveling-wave
states.}%
\label{fig:eigenmodes}%
\end{figure*}

A periodic orbit $\left(  p,n\right)  $ in the billiard is a trajectory that
closes after $p$ periods of the radial coordinate $\xi$ and $n$ periods of the
angular coordinate $\eta$ \cite{P05}.

For rotational $\mathcal{R}$-type trajectories $\left(  \gamma>0\right)  ,$
the characteristic equation for the values of $\gamma_{p,n}$ to get periodic
orbits $\left(  p,n\right)  $ is given by
\begin{equation}
\sqrt{1-\frac{f^{2}}{b^{2}}\gamma_{p,n}}=\mathrm{sn}\left[  \frac{2n}%
{p}K(\epsilon_{C}^{2})\right]  ,\qquad%
\begin{array}
[c]{l}%
\gamma_{p,n}>0,\\
p\geq3,\\
n<p/2,
\end{array}
\label{ChEqR}%
\end{equation}
where $K(\kappa)=F(\pi/2,\kappa)$ is the complete elliptic integral of the
first kind, $\mathrm{sn}(u)$ is the Jacobian elliptic sine function, and
$\epsilon_{C}=\left(  1+\gamma_{p,n}\right)  ^{-1/2}$ is the eccentricity of
the elliptic caustic. As shown in Fig. \ref{F1_fig:tray_clas}(a), $p$ is the
number of bounces at the boundary, and $n$ the number of turns around the
interfocal line in a complete cycle of the particle.

For librational $\mathcal{L}$-type trajectories $\left(  \gamma<0\right)$, the characteristic equation is
\begin{equation}
\sqrt{\frac{b^{2}}{b^{2}-f^{2}\gamma_{p,n}}}=\mathrm{sn}\left[  \frac{2n}%
{p}K\left(  \frac{1}{\epsilon_{C}^{2}}\right)  \right]  ,\qquad%
\begin{array}
[c]{l}%
\gamma_{p,n}<0,\\
p\geq4,\\
n<p/2,
\end{array}
\label{ChEqL}%
\end{equation}
where $p$ must be an even integer to have closed $\mathcal{L}$-orbits.
Furthermore, in addition to Eq. (\ref{ChEqL}), a $\mathcal{L}$-orbit $\left(p,n\right)  $ can be present in the billiard only if it satisfies the cut-off condition $\sin\left(  n\pi/p\right)  \geq b/a$. In Fig.
\ref{F1_fig:tray_clas}(b), we show some typical librational trajectories in the elliptic billiard.

Regardless of the ellipse's eccentricity, rotational and librational trajectories are always possible in the billiard. However, rotational trajectories are more likely to occur if the eccentricity is low, i.e., when the ellipse is more circular. Conversely, when the boundary is highly eccentric, librational trajectories are favored. This is why, in Figure \ref{F1_fig:tray_clas}, we have used more elongated ellipses to show the librational trajectories. 

\subsection{Quantum eigenstates}

Eigenstates $\psi\left(  \xi,\eta\right)  $ of the particle confined in the elliptic billiard are determined by solving the two-dimensional time-independent Schr\"{o}dinger equation in elliptic coordinates, namely%
\begin{equation}
\left[  \frac{\partial^{2}}{\partial\xi^{2}}+\frac{\partial^{2}}{\partial
\eta^{2}}+\frac{Mf^{2}E}{\hbar^{2}}\left(  \cosh2\xi-\cos2\eta\right)
\right]  \psi=0, \label{SEq}%
\end{equation}
where $E$ is the energy of the state $\psi\left(  \xi,\eta\right)  $ and
$\hbar$ is the reduced Planck constant. Applying the Dirichlet
condition $\psi\left(  \xi_{0},\eta\right)=0$ at the boundary, the even (e) and odd (o)
eigenfunctions are given by
\begin{align}
\psi_{r,m}^{e}(\xi,\eta)  &  =\mathcal{N}_{r,m}^{e}~\mathrm{Mc}_r^{(1)}
(\xi,q_{r,m}^{e})~\mathrm{ce}_{r}(\eta,q_{r,m}^{e}),\label{evenmode}\\
\psi_{r,m}^{o}(\xi,\eta)  &  =\mathcal{N}_{r,m}^{o}~\mathrm{Ms}_r^{(1)}
(\xi,q_{r,m}^{o})~\mathrm{se}_{r}(\eta,q_{r,m}^{o}), \label{oddmode}
\end{align}
where $\mathrm{Mc}_r^{(1)}\left(  \xi,q\right)  $ and $\mathrm{Ms}_r^{(1)}\left(\xi,q\right)$ are the even and odd radial Mathieu functions (RMF) of the first kind with integer order $r$ and parameter $q$, $\mathrm{ce}_{r}(\eta,q)$ and $\mathrm{se}_{r}(\eta,q)$ are the even and odd angular Mathieu functions (AMF) \cite{McLachlan,NISTBOOK,MeixnerBOOK}, and $\mathcal{N}_{r,m}^{e,o}$ are normalization constants such that
\begin{equation}
\left\langle \psi_{r,m}^{\sigma}|\psi_{r^{\prime},m^{\prime}}^{\sigma^{\prime }}\right\rangle =\delta_{r,r^{\prime}}\delta_{m,m^{\prime}}\delta_{\sigma,\sigma^{\prime}},\qquad\sigma=\left\{  e,o\right\}  .
\end{equation}
Eigenstates (\ref{evenmode}) and (\ref{oddmode}) represent \textit{standing wave} solutions of the Schr\"{o}dinger equation (\ref{SEq}), and form a complete orthonormal family of real solutions of the Schr\"{o}dinger equation in an elliptic domain subject to the Dirichlet condition $\psi\left(\xi_{0},\eta\right)=0$ at the boundary. 

Figure \ref{fig:eigenmodes} shows the probability distribution $\left\vert
\psi_{r,m}^{\sigma}\right\vert ^{2}$ of several even and odd eigenstates in the elliptic billiard. The pattern of the state $\psi_{r,m}^{\sigma}$ has $r$ hyperbolic nodal lines defined by the roots of the AMFs, e.g., $\mathrm{ce}_{r}(\eta,q_{r,m}^{e})=0$, and $m$ elliptic nodal lines corresponding to the roots of the RMFs, e.g., $\mathrm{Mc}_r^{(1)}(\xi,q_{r,m}^{e})=0$. Even and odd eigenstates are symmetrical and anti-symmetrical about the $x$-axis.

The energies of the eigenstates $\psi_{r,m}^{\sigma}\left(  \xi,\eta\right)$ are
\begin{equation}
E_{r,m}^{\sigma}=\left(  \frac{2\hslash^{2}}{Mf^{2}}\right)  q_{r,m}^{\sigma},
\label{quantum_energy}%
\end{equation}
where the (non-dimensional) parameter $q_{r,m}^{\sigma}$ is the $m$-th parametric zero of the $r$th-order RMF that satisfies the Dirichlet conditions at the elliptic boundary, i.e.,
\begin{equation}
\mathrm{Mc}_r^{(1)}(\xi_{0},q_{r,m}^{e})=0,\qquad \mathrm{Ms}_r^{(1)}(\xi_{0},q_{r,m}^{o})=0,
\label{BCs}
\end{equation}
depending on whether the state is even or odd. 
The term \textit{parametric zero} refers to the fact that we must calculate the value of the parameter $q$ that makes the function vanish for the value of its variable $\xi=\xi_0$.
From Eq. (\ref{quantum_energy}), it is clear that the parameter $q_{r,m}^{\sigma}$ is the energy of the eigenstate $\psi^{\sigma}_{r,m}$ in units of $2\hslash^{2}/Mf^{2}.$

The eigenstates $\psi_{r,m}^{e}$ and $\psi_{r,m}^{o}$ are not degenerate for
the same indices $(r,m)$. As $q$ decreases, eigenvalues of the even and odd
eigenstates get closer to each other. They are equal only in the limiting
case when $q=0$, i.e., when the elliptic boundary reduces to a circle.

The eigenvalues of the second constant of motion $\Gamma$ can be obtained by
substituting the momentum operators $p_{\xi}=-\mathrm{i}\hslash\partial
/\partial\xi$ and $p_{\eta}=-\mathrm{i}\hslash\partial/\partial\eta$ in Eq.
(\ref{LambdaEquation}), we get%
\begin{equation}
\frac{f^{2}\hslash^{2}}{h_{\bot}^{2}}\left(  \sin^{2}\eta\frac{\partial^{2}}%
{\partial\xi^{2}}-\sinh^{2}\xi\frac{\partial^{2}}{\partial\eta^{2}}\right)
\psi=\Gamma\psi.
\end{equation}
Applying the equivalence between the eigenvalue problem for $\Gamma$ and the
eigenvalue problem for the Hamiltonian Eq. (\ref{SEq}), the eigenvalues of the
normalized constant of motion $\gamma$ [Eq. (\ref{gama})] can be calculated with
\begin{equation}
\gamma_{r,m}^{\sigma}=\frac{\alpha_{r,m}^{\sigma}-2q_{r,m}^{\sigma}}%
{4q_{r,m}^{\sigma}}, \label{quantumgamma}%
\end{equation}
where $\alpha_{r,m}^{\sigma}$ is the $r$-th characteristic value of the angular
Mathieu function $\mathrm{ce}_{r}(\eta,q_{r,m}^{e})$ or $\mathrm{se}_{r}%
(\eta,q_{r,m}^{o})$ depending on the parity $\sigma=\left\{e,o\right\}$ \cite{McLachlan}.

In analogy with the classical mechanics solution, positive values of
$\gamma_{r,m}^{\sigma}$ are associated with \textit{rotational} $\mathcal{R}%
$-type eigenstates, while negative values of $\gamma_{r,m}^{\sigma}$ are
associated with \textit{librational} $\mathcal{L}$-type eigenstates. The
separatrix between both kinds of motion is given by the straight line
$\alpha=2q$ on the $\left(  \alpha,q\right)  $ plane. Librational states are
always non-degenerate, but $\mathcal{R}$-type states become more degenerate as
$\gamma$ increases.%

Alternatively, \textit{traveling-wave} complex eigenstates of the elliptic
billiard can be constructed by the linear superposition of the even and odd
standing-wave states%
\begin{equation}
\psi_{r,m}^{\pm}(\xi,\eta)=\frac{1}{\sqrt{2}}\left[  \psi_{r,m}^{e}(\xi
,\eta)\pm\mathrm{i}\psi_{r,m}^{o}(\xi,\eta)\right]  . \label{TW}%
\end{equation}
As shown in Fig. \ref{fig:eigenmodes}(c), the probability density pattern
$\left\vert \psi_{r,m}^{+}\right\vert ^{2}$ of the traveling states has an
elliptic ringed structure. The corresponding phase distributions$\ \arg
\psi_{r,m}^{+}$ are illustrated in Fig. \ref{fig:eigenmodes}(d). For the
traveling states with $r=1$, the phase exhibits a single vortex at the origin.
For $r\geq2$, the phase patterns has $r$ in-line vortices, each with unitary
topological charge such that the total charge (along a closed trajectory
enclosing all the vortices) is $r$. The branch cuts lie upon confocal
hyperbolas, implying that, on time evolution, a point in the phase
distribution travels along an elliptic path of constant $\xi$. Note that the
quantum probability current $\mathbf{J}\propto\mathrm{Im}(\psi^{\ast}%
\nabla\psi)$ rotates around the interfocal line of the billiard. Since
$\psi_{r,m}^{+}$ and $\psi_{r,m}^{-}$ are symmetrical in spatial structure,
only the case of $\psi_{r,m}^{+}$ is shown.

When the elliptic billiard becomes a circular billiard, the traveling wave
solutions $\psi_{r,m}^{\pm}(\xi,\eta)$ reduce to the known solutions
$J_{r}\left(  \kappa_{r,m}\rho\right)  e^{\pm\mathrm{i}r\theta}$ in circular
polar coordinates $\left(  \rho,\theta\right)$. In this case, the $r$
in-line vortices of the elliptic solution degenerate into a single high-order
vortex at the origin with charge $r$.

\section{Quantum states localized on the classical periodic orbits}

\subsection{Quantum-classical connection}

It is clear that if the potential energy is zero inside the billiard, the form of a classical periodic trajectory is not affected by a change in the (kinetic) energy $E$ of the particle. Thus, regardless of its energy, a periodic classical trajectory $(p,n)$ is
characterized entirely by the value of the parameter $\gamma_{p.n}$ that
satisfies one of the characteristic equations (\ref{ChEqR}) or (\ref{ChEqL})
depending on whether the trajectory is $\mathcal{R}$-type or $\mathcal{L}$-type. 
On the other hand, in the quantum description, an eigenstate $\psi_{r,m}^{\sigma}(\xi,\eta)$ has associated a specific eigenenergy $E_{r,m}^{\sigma}$ and a parameter $\gamma_{r,m}^{\sigma}$. The values of these two conserved quantities are not independent but are related by Eqs. (\ref{quantum_energy}) and (\ref{quantumgamma}).

A quantum-classical connection in the elliptic billiard can be established
using Eq. (\ref{quantumgamma}) by substituting the parameter $\gamma_{p,n}$
corresponding to the classical periodic orbit $(p,n).$ Solving for $q,$ we
introduce the following semi-classical parameter:
\begin{equation}
q_{\mathrm{sc}}(p,n,r_{0},m_{0})=\frac{\alpha^e_{r_0,m_0}}{2+4\gamma_{p,n}},
\end{equation}
where $\left(  r_{0},m_{0}\right)  $ are the indices of the central eigenstate
$\psi_{r_{0},m_{0}}^{\pm}(\xi,\eta)$ around which we will build the
superposition of nearly degenerate eigenstates.

\begin{figure}[t]
\centering
\includegraphics{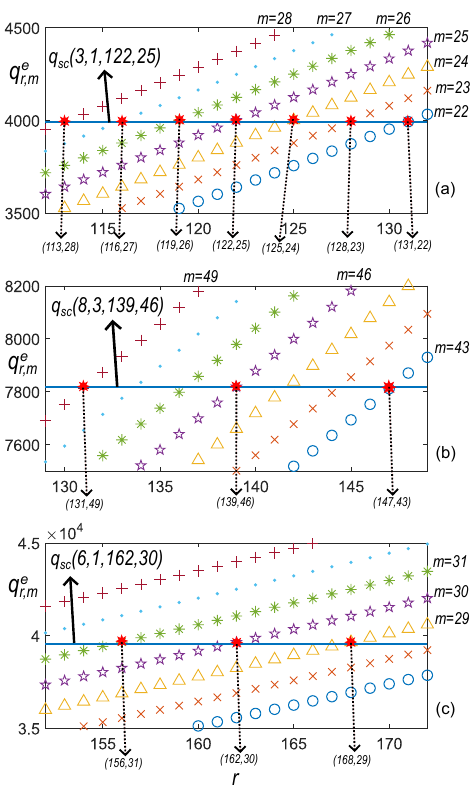}
\caption{Characteristic values of $q_{r,m}^{e}$ in function of the order $r$
for a billiard with eccentricity (a) (b) $\epsilon=0.5,$ and (c)
$\epsilon=0.924.$ In each subplot, the horizontal solid line corresponds to
the value of $q_{\mathrm{sc}}(p,n,r_{0},m_{0})$ and each marker defines a
particular $m.$ The values $q_{r,m}^{e}$ closest to the horizontal line
correspond to the degenerate states.}%
\label{fig3:degenerancy_cond}%
\end{figure}
\begin{table}[b]
\caption{\label{tab:exact_val}%
Exact values of $q_{\mathrm{sc}}$, $q_{nm}$, $\gamma_{p,n}$ and $\gamma
_{r_{0},m_{0}}$ from Fig. \ref{fig3:degenerancy_cond} . It can be seen that
$q_{r_{0},m_{0}} \approx q_{sc}$ and $\gamma_{p,n} \approx\gamma_{r_{0},m_{0}%
}$. The high value of the case $\mathcal{L}$(6,1) comes from considering a different
ellipse geometry.
}
\begin{ruledtabular}
\begin{tabular}{lrrrr}
$(p,n)$ & $q_{sc}$ & $q_{r_{0},m_{0}}$ & $\gamma_{p,n}$ & $\gamma_{r_{0},m_{0}}$
\\
\colrule
$\mathcal{R}$(3,1) & 4002.23 & 3991.42 & 0.4667 & 0.4641\\
$\mathcal{R}$(8,3) & 7819.11 & 7819.16 & 0.1714 & 0.1714\\
$\mathcal{L}$(6,1) & 39577.28 & 39645.37 & -0.2786 & -0.2790\\
\end{tabular}
\end{ruledtabular}
\end{table}

Constructing POQS related to classical periodic orbits is
notationally facilitated by the traveling wave representation Eq. (\ref{TW}). Every
eigenstate in the billiard is characterized by a set of quantum numbers
$(r,m)$, a parameter $q_{r,m}$, a characteristic value $\alpha_{r,m},$ and a
parameter $\gamma_{r,m}$. By giving a central order $r_{0}$ and a classical
parameter $\gamma_{p,n}$, we can find $m_{0}$ that makes $q_{\mathrm{sc}}$ and
$q_{r_{0},m_{0}}$ as equal as possible. This is equivalent to finding $m_{0}$
that makes $\gamma_{p,n}$ and $\gamma_{r_{0},m_{0}}$ as similar to each other
as possible. Both scenarios are achieved at the same $m_{0}$. The energies and
the parameters $\gamma_{r,m}$, $q_{r,m}$ are slightly different in the even
and odd states for the same quantum numbers, especially in the $\mathcal{L}%
$-type motion \cite{P05}. 
Thus, we adopted the eigenvalues of the real part of the traveling waves $\psi_{r_{0},m_{0}}^{+}$ to accomplish $q_{r_{0},m_{0}}\approx q_{\mathrm{sc}}$ or $\gamma_{p,n}\approx\gamma_{r_{0},m_{0}}$.

In Fig. \ref{fig3:degenerancy_cond}, we plot the values of $q^e_{r,m}$ for different quantum numbers $(r,m)$ associated with the periodic orbits $\mathcal{R}(3,1)$, $\mathcal{R}(8,3)$, and $\mathcal{L}(6,1)$. Each marker represents a particular number $m$. The horizontal lines correspond to the value of $q_{\mathrm{sc}}$ for each classic orbits $(p,n)$ and different central orders $(r_{0},m_{0})$. In Table I, we include the explicit values of $q_{\mathrm{sc}},q_{r_{0},m_{0}},\gamma_{p,n},\gamma_{r_{0},m_{0}}$ for each of the classic periodic orbits considered in Fig. \ref{fig3:degenerancy_cond}. Values closer to the horizontal lines are used to build the nearly degenerate states. Empirically, the quantum numbers $(r,m)$ of the superposing eigenstates can be found with%
\begin{equation}
r(s)=r_{0}\pm sp,\qquad m(s)=m_{0}\mp sn, 
\label{DDC}%
\end{equation}
where $s=0,1,2,...,S$. The maximum value $S$ is determined by specifying a maximum tolerance for the difference $|q_{\mathrm{sc}}-q_{r,m}|$. 

Similar expressions to Eqs. (\ref{DDC}) were obtained for the circular billiard with eigenstates $J_{r}\left(  \kappa_{r,m}\rho\right)  e^{\pm\mathrm{i}r\theta}$ \cite{Liu2006,hsieh2017extracting}. But in this case, the energy depends on the square of the parameter $\kappa_{r,m}$, which represents a zero of the Bessel function $J_{r}$, while in the elliptic billiard, the energy is proportional to $q$ [Eq. (\ref{quantum_energy})]. For this reason, in the elliptic case, we deal with higher values of the zeros of the RMFs. This could imply that the value of $S$ may be higher than in the case of the circular billiard. As we will see later, we found that a value of $S=2$ seems sufficient for assembling POQS localizing classical periodic orbits in the elliptic billiard.

\subsection{Numerical aspects}

It is worth commenting on the numerical aspects involved in computing the radial Mathieu functions $\mathrm{Mc}_r^{(1)}(\xi,q)$ and $\mathrm{Ms}_r^{(1)}(\xi,q)$ and their parametric zeros. Several algorithms are available for computing the RMFs \cite{Leeb,Alhargan2000,Alhargan2006,Erricolo,Cojocaru}, but we found that they provide acceptable accuracy only for moderate values of $q$, typically less than 300, and low orders, i.e., $r<100$. However, to establish a good connection between the classical trajectories and quantum states, we need to evaluate RMFs with higher orders $(r>150)$ and very large $q$, i.e., $(q>10,000)$, as illustrated in the axes of Fig. \ref{fig3:degenerancy_cond}. 

To achieve these requirements, we developed our own computational routines to evaluate the RMFs and accurately calculate their parametric zeros. First, to compute the characteristic values $\alpha$ of the Mathieu functions, we implemented an efficient matrix method that guarantees a precision of about $10^{-12}$ for the orders $r$ and values $q$ required in this work. Once the characteristic values were calculated, we computed the RMFs by evaluating their expansions in terms of products of modified Bessel functions \cite{NISTBOOK,MeixnerBOOK} using a novel adaptive method \cite{VanBuren}. 

We tested our algorithms by assessing the analytical Wronskians of the RMFs and the plane wave expansions in terms of Mathieu waves \cite{McLachlan,P02}, giving an accuracy of about $10^{-10}$. Once we developed reliable routines to evaluate the RMFs in terms of its order $r$, the argument $z$, and the parameter $q$, we applied an iterative Newton-Raphson method to compute the parametric zeros $q_{r,m}$ of the even and odd Mathieu functions to satisfy the Dirichlet boundary conditions Eqs. (\ref{BCs}).

\subsection{Construction of the periodic-orbit quantum states}

After applying the degenerate condition Eq. (\ref{DDC}), the wave function associated with the classical periodic orbit $(p,n)$ in the elliptic billiard is given by the superposition of nearly degenerate traveling states $\psi_{r,m}^{+}(\xi,\eta)$, namely
\begin{multline}
\Psi_{S,r_{0},m_{0}}^{p,n}(\xi,\eta;\phi_{0})=
\label{coherentstate}\\
\frac{1}{2^S} \sum_{s=-S}^{S}
\begin{pmatrix}
2S\\
S+s
\end{pmatrix}
^{1/2}e^{\pm\mathrm{i}s\phi_{0}}\psi_{r_{0}+sp,m_{0}-sn}^{+}(\xi,\eta),
\end{multline}

where the expansion coefficients have an amplitude binomial factor and an exponential phase factor $e^{\pm \mathrm{i} s\phi_{0}}$. 

The binomial amplitude factor was derived originally from expanding generalized coherent modes in terms of the angular momentum states of the harmonic oscillator, as shown by Wodkiewicz and Eberly \cite{wodkiewicz1985coherent}, among others \cite{buvzek1989generalized,de1992oscillator,Chen2005}. Later, the same amplitude factor was applied successfully in the expansion of generalized states in square, triangular, and circular billiards \cite{Chen2002,chen2003vortex,Liu2006}. Since the elliptical billiard is a generalization of the circular one, we chose this factor as it can also be applied in the limiting case. As we will show below, this selection was appropriate. 
The overall factor $2^{-S}$ ensures that the function $\Psi_{S,r_{0},m_{0}}^{p,n}$ is normalized, provided that the constituent wavefunctions $\psi_{r,m}^{+}(\xi,\eta)$ are normalized,  as is indeed the case.

The phase factor $e^{\pm \mathrm{i} s\phi_{0}}$ controls the relative phase between the constituent eigenstates. As verified numerically, this factor is related to the initial conditions of the associated classical periodic orbit and plays an important role in the quantum-classical connection, which has also been confirmed in the studies of the harmonic oscillator \cite{wodkiewicz1985coherent,buvzek1989generalized,de1992oscillator,Chen2005}. The positive and negative signs in the phase factor represent a clockwise or counterclockwise rotation of the associated classical trajectory. An independent \textit{classical} state can be constructed by superposing the wavefunctions with negative helicity, i.e., $\psi_{r,m}^{-}\left(  \xi,\eta\right),$ but the final effect is only changing the sign of the imaginary part of $\Psi_{S,r_{0},m_{0}}^{p,n}$.

\begin{figure}[t]
\centering
\includegraphics{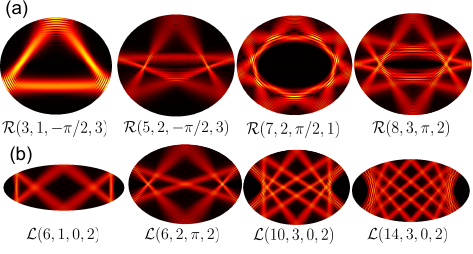}
\caption{POQS with localization on the classical periodic orbits
shown in Fig. \ref{F1_fig:tray_clas} for some combinations of the parameters
$(p,n,\phi_{0},S)$ using $r_{0}=50$ for $\mathcal{R}(3,1,3\pi/2,3)$,
$r_{0}=120$ for $\mathcal{R}(5,2,3\pi/2,3),$ $\mathcal{R}(7,2,\pi/2,1),$
$\mathcal{R}(8,3,\pi,2)$ and $r_{0}=130$ for $\mathcal{L}(6,1,0,2),$
$\mathcal{L}(6,2,\pi,2),$ $\mathcal{L}(10,3,0,2),$ and $\mathcal{L}%
(14,3,0,2).$}%
\label{fig4:tray_quantum}%
\end{figure}

In Fig. \ref{fig4:tray_quantum}, we show the probability density of the POQS $|\Psi_{S,r_{0},m_{0}}^{p,n}|^2$ corresponding to the $\mathcal{R}$-type and
$\mathcal{L}$-type classical trajectories depicted in Fig. \ref{F1_fig:tray_clas}. The similarity between the localized POQS and the classical orbits is evident. The values of the parameters used in the superposition are included in the figure caption. In regions near the boundary where the trajectories bounce, the incident wave superposes with the reflected one, resulting in interference fringes of probability. This phenomenon can also be observed where the trajectories self-intersect within the billiard. The triangular trajectory (3,1) in Fig. \ref{fig4:tray_quantum}(a) contains two diagonal segments that exhibit the focusing of the probability density. When we examine only one of these segments, the probability density resembles the Gaussian intensity distribution between two spherical mirrors in a laser cavity. We will revisit this effect when we discuss phases and probability current later.

Single eigenstates $\psi_{r,m}^{\sigma}$ of the elliptic billiard do not resemble the classical periodic orbits, even considering large quantum numbers $\left(r,m\right)$. However, even if only three nearly degenerate eigenstates ($S=1$) are properly superimposed, the resulting POQS could localize the classical periodic orbit very accurately.

\begin{figure}[t]
\centering
\includegraphics{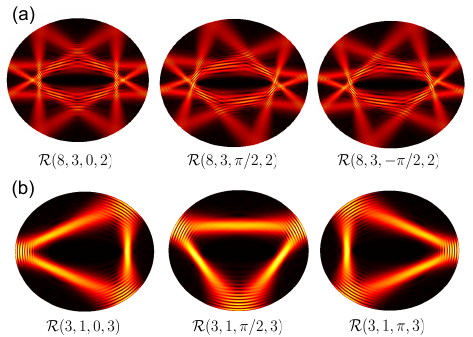}
\caption{Effect of the phase factor $\phi_{0}$ in the $\mathcal{R}$-type
POQS. (a) State $\mathcal{R}(8,3,\phi_{0},2)$. (b) State
$\mathcal{R}(3,1,\phi_{0},3)$. Central quantum numbers $(r_{0},m_{0})$ are the
same as Fig. \ref{fig4:tray_quantum} for the corresponding trajectories.}%
\label{fig:efecto_phi0_R}%
\end{figure}

For $\mathcal{R}$-type POQS [Fig. \ref{fig4:tray_quantum}(a)], the effect of the phase factor $\phi_{0}$ is to displace the position of the impact points of the localized trajectories on the elliptic boundary.  This result was confirmed numerically, and it is illustrated in Fig. \ref{fig:efecto_phi0_R} where we plot the wave patterns of the rotational trajectories $\mathcal{R}(8,3,\phi_{0},2)$ and $\mathcal{R}(3,1,\phi_{0},3)$ for different values of
$\phi_{0}$. 
By adjusting the value of $\phi_{0}$, we can start the trajectory at a specific point on the boundary, making the classic trajectory symmetric or asymmetric with respect to the Cartesian axes. The continuous variation of $\phi_{0}$  would correspond to a kind of rotation of the localized classical trajectory within the billiard. Because the
elliptic boundary does not have rotational symmetry about the origin, starting
the orbit at a different point on the boundary leads the trajectory $(p,n)$ to have different shapes. However, regardless of the $\phi_{0}$ value, all the displaced
orbits share the same conserved quantities. This effect has also been observed
in the triangular and square billiards \cite{chen2003vortex,chen2002quantum_square3}. In the case of the circular billiard, because of the
angular symmetry of the boundary, changing the phase factor $\phi_{0}$ only produces a
trivial rotation of the same pattern about the origin \cite{Liu2006}.%

\begin{figure}[t]
\centering
\includegraphics{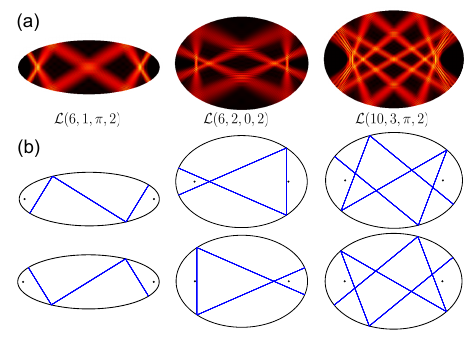}
\caption{Effect of the phase factor $\phi_{0}$ in the $\mathcal{L}$-type
POQS. (a) States $\mathcal{L}(6,1,\pi,2)$, $\mathcal{L}(6,2,0,2),$
and $\mathcal{L}(10,3,\pi,2)$. (b) Decomposition in primitive periodic
orbits.}%
\label{fig:efecto_phi0_L}%
\end{figure}

For $\mathcal{L}$-type POQS [Fig. \ref{fig4:tray_quantum}(b)], the variation of $\phi
_{0}$ also shifts the bouncing points of the orbits on the boundary, as occurs
in $\mathcal{R}$-type trajectories. However, the phenomenology of librational
$\mathcal{L}$-type states is more complex than $\mathcal{R}$-type ones. 
In particular, there is the possibility that, for specific values of $\phi_{0},$ the periodic closed trajectories split into two or more primitive trajectories, as shown in Fig. \ref{fig:efecto_phi0_L}, where we plot the same states as in Fig. \ref{fig4:tray_quantum}(b) but with different phase factors $\phi_{0}$. 
In the cases shown, the pairs of primitive orbits are symmetric about the $y$-axis.
Note that each primitive trajectory in Fig. \ref{fig:efecto_phi0_L} may seem to have a starting and an ending point, but actually, the particle is bouncing perpendicularly off the boundary and returning along the same path.
This process is happening repeatedly, and the superposition of quantum states naturally reconstructs both degenerate primitive trajectories. It is expected that for noncoprime indices $(r,n)$, the wave function of the POQS would be localized on multiple periodic orbits, as it was shown in previous studies of the harmonic oscillator and the triangular and circular billiards \cite{Chen2003HO,makowski2005comment,hsieh2017extracting}. But in the case of the $\mathcal{L}$-type states in the elliptic billiard, the POQS is naturally composed of multiple periodic orbits.%

\begin{figure}[t]
\centering
\includegraphics{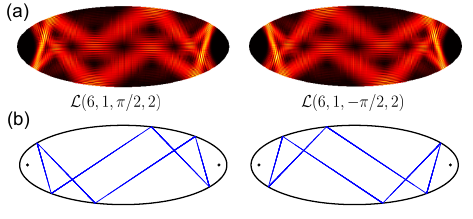}
\caption{(a) $\mathcal{L}$-type POQS with a phase shift of $\pi$.
(b) Classical symmetric trajectories associated with quantum states plotted in (a).}%
\label{fig:61_tray_quantum_90_270}%
\end{figure}

To further confirm that the $\mathcal{L}$-type POQS are composed of
two symmetric independent classical trajectories, Fig.
\ref{fig:61_tray_quantum_90_270} shows the POQS $\mathcal{L}%
(6,1,\pi/2,2)$ and $\mathcal{L}(6,1,-\pi/2,2)$ and their corresponding
classical orbits. It seems that both POQS are different, although
symmetrical. However, upon closer inspection, we notice the existence of regions
where the state's amplitude is very low, but it still exists. With this in
mind, we can conclude that both POQS are constructed using the same
symmetrical trajectories.%

\begin{figure}[t]
\centering
\includegraphics{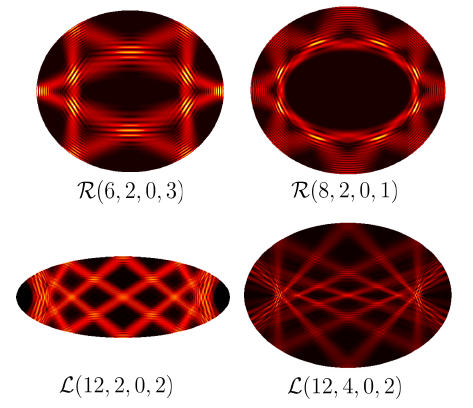}
\caption{Rotational and librational POQS with noncoprime indices
$(p,n)$.}%
\label{fignoncoprime}%
\end{figure}

Finally, the case of noncoprime indices $(p,n)$ is shown in Fig.
\ref{fignoncoprime}. It can be seen that the $\mathcal{R}$-type state
$\mathcal{R}(6,2,0,3)$ splits into two independent primitive trajectories
$\mathcal{R}(3,1)$ with different initial conditions. These POQS
are composed of two individual states with a phase difference $\phi_{0}$ of
$2\pi/j$ where $j$ is the common factor between the indices $(p,n),$ and
therefore, $j$ also gives the number of independent trajectories that compose
the $\mathcal{R}$-type POQS. In this way, the state $\mathcal{R}%
(8,2,0,1)$ splits into two $\mathcal{R}\left(  4,1\right)  $ states. In the
case of $\mathcal{L}$-type trajectories, because the POQS is
composed of two independent trajectories, the POQS with $j=2$ [see
$\mathcal{L}(12,2,0,2)$ and $\mathcal{L}(12,4,0,2)$ in Fig.
\ref{fignoncoprime}] has four independent trajectories, given by
$\mathcal{L}(6,1,0,2)$, $\mathcal{L}(6,1,\pi,2)$ for $\mathcal{L}(12,2,0,2)$
and $\mathcal{L}(6,2,0,2)$, $\mathcal{L}(6,2,\pi,2)$ for $\mathcal{L}%
(12,2,0,2)$. Thus, for the $\mathcal{L}$-type state with noncoprime indices
$(p,n)$, the number of independent classical trajectories are $2j$, and they
are composed of $j$ individual constituent states with a phase difference of
$2\pi/j$ between them as occurred for the $\mathcal{R}$-type states with
noncoprime indices $(p,n)$.%

\subsection{Probability current density and phase distribution}

For a wavefunction in the billiard $\Psi=\left\vert \Psi\right\vert \exp(\mathrm{i}\beta)$, the probability current density $\mathbf{J}$ is given by
\begin{equation}
\mathbf{J}(\xi,\eta;\phi_{0}) = \frac{\hbar}{M} \mathrm{Im}(\Psi^{\ast}\nabla \Psi) = \frac{\hbar}{M} \left\vert \Psi\right\vert ^{2}~\nabla\beta,
\label{J}
\end{equation}
where $\beta = \beta(\xi,\eta)$ is the phase distribution of the wavefunction whose gradient determines the direction of the probability flow on the surface of the billiard.

\begin{figure}[t]
\centering
\includegraphics[width=7cm]{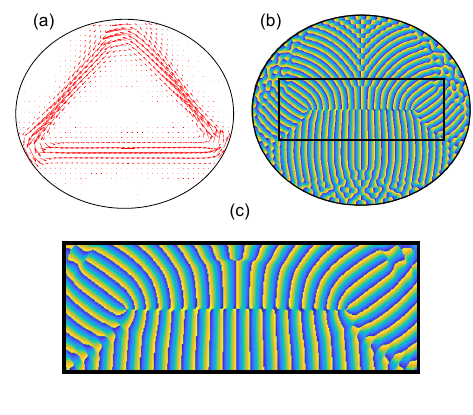}
\caption{(a) Quantum probability current for the POQS $\mathcal{R}%
(3,1,-\pi/2,3)$ with $r_0=50$. (b) Phase distribution. (c) Zoom of the phase around the
interfocal line.}%
\label{fig:quantum flux 31}%
\end{figure}

Figure \ref{fig:quantum flux 31} shows the probability current and the phase distribution of the localized trajectory $\mathcal{R}(3,1)$ depicted in Fig. \ref{fig4:tray_quantum}(a). 
The probability flows counterclockwise inside the billiard, reflecting on the walls, creating a closed periodic orbit. The direction of the probability flow can be reversed by simply conjugating $\Psi$. The size of the probability vectors is larger in regions with higher probability density.  

\begin{figure}[t]
\centering
\includegraphics[width=7cm]{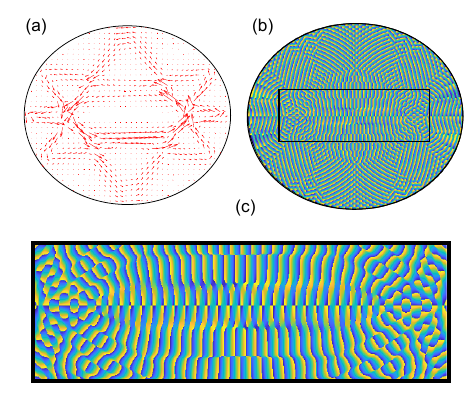}
\caption{(a) Quantum probability current for the POQS $\mathcal{R}%
(8,3,\pi,2)$ with $r_{0}=70$. (b) Phase distribution. (c) Zoom of the phase
around the interfocal line.}%
\label{fig:quantum flux 83}%
\end{figure}

The phase distribution exhibits a complicated structure of vortices throughout the area enclosed by the elliptic wall. A vortex appears at a zero-probability point where the real and the imaginary parts of the complex function $\Psi$ vanish, and thus, the phase there is not defined. All vortices have a topological charge equal to one. Note that the probability current vector circulates the vortices. The line of vortices that appears along the $x$-axis at the interfocal line is particularly interesting. To better appreciate it, in Fig. \ref{fig:quantum flux 31}(c), we show a zoom of that region. The number of interfocal vortices is proportional to the order $r_0$ of the central eigenstate of the superposition. This result is predictable since the angular Mathieu function of order $r$ has $r/2$ zeros in the interval $[0,\pi]$ \cite{McLachlan}, which is a necessary condition for the state to vanish at those points.

Observing the phase wavefronts in Figure \ref{fig:quantum flux 31}(b) also provides valuable insights. If we focus our attention on the region through which one of the diagonal segments of the trajectory passes, we observe that the wavefronts resemble the typical converging and diverging spherical wavefronts of a Gaussian beam in a cavity composed of two spherical mirrors. The plane of maximum beam focusing (in our case, maximum probability) corresponds to the plane wavefront with zero curvature. These results show the close analogy between wave propagation in optical cavities and quantum distributions in mesoscopic billiards.

In Fig. \ref{fig:quantum flux 83}, we show the probability current and phase distribution of the localized trajectory $\mathcal{R}(8,3)$ depicted in Fig. \ref{fig4:tray_quantum}(a). We include this example to illustrate the case of self-intersecting trajectories, which occur when $n>1$. 
For the $\mathcal{R}$-type orbits, the vector current coincides with the classical velocities throughout the closed orbit. We can follow the direction of the vectors along the entire closed trajectory in the same way as we would with the classical trajectory $\mathcal{R}(8,3)$ shown in Fig. \ref{F1_fig:tray_clas}(a).

\begin{figure}[t]
\centering
\includegraphics[width=5cm]{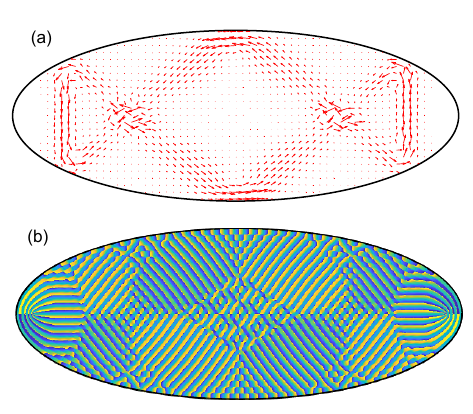}
\caption{(a) Quantum probability current for the POQS $\mathcal{L}%
(6,1,0,2)$ with $r_{0}=70$. (b) Phase distribution.}%
\label{fig:quantum flux 61}%
\end{figure}

Figure \ref{fig:quantum flux 61} shows the case of an $\mathcal{L}(6,1)$ librational trajectory. In contrast to the rotational case, the phase of $\mathcal{L}$-type orbits does not display a well-defined series of vortices along the interfocal line. This is because librational trajectories always cross the $x$-axis within the foci of the boundary. If we follow the probability vectors along the entire closed trajectory, we see that their direction in the vertical segments appears reversed compared to the direction expected in the velocity vector of a classical orbit. We attribute this discrepancy to the fact that $\mathcal{L}$-type can be divided into primitive trajectories, and a sign change is introduced in the superposition.

\section{Integral line equations extracted from the POQS}

Let us now show how the classical periodic trajectories can be extracted from the POQS $\Psi_{S,r_{0},m_{0}}^{p,n}(\xi,\eta;\phi_{0})$ in the elliptic billiard. In Eq. (\ref{coherentstate}), the amplitude factor could be written as $Ae^{\pm\mathrm{i}\phi_{0}},$ but, according to Ref. \cite{inomata1992path}, when the amplitude $A$ is equal to unity, it is reached the minimum uncertainty for the superposed state. In view of this, we first modify Eq. (\ref{coherentstate}) by setting the weight for each eigenstate to be unity and multiplying by $p$ to the phase factor; we get
\begin{multline}
\widetilde{\Psi}_{S,r_{0},m_{0}}^{p,n}(\xi,\eta;\phi_{0})=\label{coherentstate_mod}\\
\frac{1}{\sqrt{2S+1}}\sum_{s=-S}^{S}e^{\mathrm{i}sp\phi_{0}}\psi
_{r_{0}+sp,m_{0}-sn}^{+}(\xi,\eta),
\end{multline}
where the tilde means that $\widetilde{\Psi}$ is an approximation of $\Psi$ in Eq. (\ref{coherentstate}).

The plane wave expansions of the even [Eq. (\ref{evenmode})] and
odd [Eq. (\ref{oddmode})] eigenstates in the elliptic billiard are \cite{McLachlan}
\begin{widetext}
\begin{subequations}
\label{PWEx}
\begin{align}
\psi_{2n,m}^{e}(\xi,\eta) &  = \mathcal{N}_{2n,m}^{e} \frac{\left(  -1\right)  ^{n}}{2\pi}\int_{-\pi}^{\pi}\mathrm{ce}_{2n}\left(  \phi,q_{2n,m}^{e}\right)  \mathbb{P}^e\mathrm{d}\phi,\label{evenmode2}\\
\psi_{2n+1,m}^{e}(\xi,\eta) &  = \mathcal{N}_{2n+1,m}^{e} \frac{\mathrm{i}\left(-1\right)^{n+1}}{2\pi}\int_{-\pi}^{\pi}\mathrm{ce}_{2n+1}\left(  \phi,q_{2n+1,m}^{e}\right) \mathbb{P}^e\mathrm{d}\phi,\\
\psi_{2n+2,m}^{o}(\xi,\eta) &  = \mathcal{N}_{2n+2,m}^{o} \frac{\left(  -1\right)  ^{n+1}}{2\pi} \int_{-\pi}^{\pi}\mathrm{se}_{2n+2}\left(  \phi,q_{2n+2,m}^{o}\right)\mathbb{P}^o\mathrm{d}\phi,\\
\psi_{2n+1,m}^{o}(\xi,\eta) &  = \mathcal{N}_{2n+1,m}^{o} \frac{\mathrm{i}\left(  -1\right)  ^{n+1}}{2\pi}\int_{-\pi}^{\pi}\mathrm{se}_{2n+1}\left(  \phi,q_{2n+1,m}^{o}\right)\mathbb{P}^o\mathrm{d}\phi.
\end{align}
\end{subequations}
\end{widetext}
where the AMFs $\mathrm{ce}_r(\phi,q)$ and $\mathrm{se}_r(\phi,q)$ are the angular spectra, and
\begin{equation}
\mathbb{P}^{\sigma} \equiv \exp \left[\mathrm{i}k^{\sigma}\rho\cos(\phi-\theta)\right], \qquad \sigma=\left\{e,o\right\}, 
\end{equation}
is a plane wave with wavenumber $k^{\sigma}=2 (q^{\sigma})^{1/2}/f$ traveling into $\phi$
direction, and $\left(  \rho,\theta\right)  $ are the polar coordinates, i.e.,
$\left(  x,y\right)  =\left(  \rho\cos\theta,\rho\sin\theta\right)  $.

By replacing these expansions into Eq. (\ref{coherentstate_mod}) and doing the
change of variable $\varphi=\phi-\phi_{0},$ we get%
\begin{multline}
\widetilde{\Psi}_{S,r_{0},m_{0}}^{p,n}(\xi,\eta;\phi_{0})=\label{line_eq1}\\
C\left[  \int_{-\pi}^{\pi}\mathrm{ce}_{r_{0}}(\varphi+\phi_{0},q^e_{r_{0},m_{0}%
})\mathbb{P}^e_{\phi_{0}}\mathcal{~D}(p\varphi)\mathrm{d}\varphi\right.  \\
+\mathrm{i}\left.  \int_{-\pi}^{\pi}\mathrm{se}_{r_{0}}(\varphi+\phi
_{0},q^o_{r_{0},m_{0}})\mathbb{P}^o_{\phi_{0}}\mathcal{~D}(p\varphi)\mathrm{d}%
\varphi\right],
\end{multline}
where $C$ is an overall normalization constant,%
\begin{equation}
\mathbb{P}^{\sigma}_{\phi_{0}}\equiv \exp \left[\mathrm{i}k^{\sigma}_{r_{0},m_{0}}\rho\cos(\varphi
+\phi_{0}-\theta)\right], 
\end{equation}
and
\begin{equation}
\mathcal{D}(p\varphi)=\frac{1}{2S+1}\sum_{s=-S}^{S}e^{-\mathrm{i}sp\varphi},
\end{equation}
is the normalized Dirichlet kernel \cite{levi1974geometric}. It turns out that $\mathcal{D}(p\varphi)$
is a periodic pulse function with period $2\pi/p$, so the integrals in Eq.
(\ref{line_eq1}) can be split into $p$ segments with the integration interval
from $-\pi/p$ to $\pi/p$. Also, for $(2S+1)p\gg1$, $\mathcal{D}(p\varphi)$ has
a narrow peak in a small region $-\Delta\leq\varphi\leq\Delta$ with
$\Delta=\pi/p(2S+1)$ and we can approximate $\mathcal{D}(p\varphi)$ to unity
in the interval $[-\Delta,\Delta]$ and zero elsewhere. Finally, the integrals
in Eq. (\ref{line_eq1}) can be expressed as
\begin{multline}
\widetilde{\Psi}_{S,r_{0},m_{0}}^{p,n}(\xi,\eta;\phi_{0}) = \label{line_eq2}\\
C\sum_{l=1}^{p}\left[  \int_{-\Delta}^{\Delta}\mathrm{ce}_{r_{0}}(\varphi+\phi_{0}+v(l), q^e_{r_{0},m_{0}})\mathbb{P}^e_{l}\mathcal{~}\mathrm{d}\varphi\right. \\
+\left.  \mathrm{i}\int_{-\Delta}^{\Delta}\mathrm{se}_{r_{0}}(\varphi+\phi_{0}+v(l),q^o_{r_{0},m_{0}}) \mathbb{P}^o_{l}\mathcal{~}\mathrm{d}\varphi\right],
\end{multline}
where
\begin{equation}
\mathbb{P}^{\sigma}_{l}\equiv \exp\left[\mathrm{i}k^{\sigma}_{r_{0},m_{0}}\rho\cos(\varphi+\phi
_{0}-\theta+v(l)) \right],
\end{equation}
and $v(l)$ is directly related to the classical velocity by%
\begin{equation}
v(l)=\arctan(p_{y_{l}}/p_{x_{l}}),
\end{equation}
with $p_{y_{l}}$ and $p_{x_{l}}$ being the momentum components for each line
segment $l=\left\{  1,2,..,p\right\}  $ belonging to the corresponding
classical trajectory.

\begin{figure}[t]
\centering
\includegraphics[width=8cm]{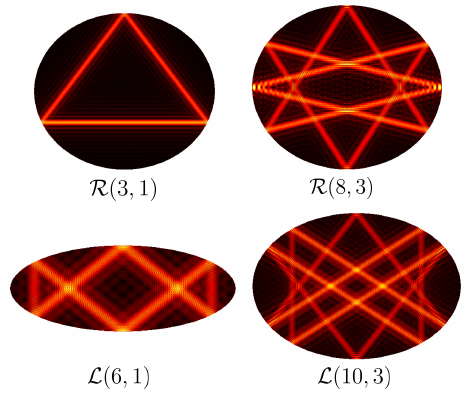}
\caption{POQS calculated with the approximated Eq. (\ref{line_eq2})
and the corresponding classical velocity components $v(l)$ for some classical
periodic orbits $(p,n)$. 
Orbit $\mathcal{R}(3,1)$ with $r_{0}=300$ and $S=5$. 
Orbit $\mathcal{R}(8,3)$ with $r_{0}=300$ and $S=2$. 
Orbit $\mathcal{L}(6,1)$ with $r_{0}=300$ and $S=6$. 
Orbit $\mathcal{L}(10,3)$ with $r_{0}=700$ and $S=5$. }
\label{fig:line_equations_figure}%
\end{figure}

The field given by Eq. (\ref{line_eq2}) localizes sharply a classical periodic trajectory with $p$ bounces. Each integral represents a line segment whose slope is controlled
by the phase factor $\phi_{0}$ and $v(l)$. In Fig. \ref{fig:line_equations_figure} we plot $|\Psi_{S,r(0),m(0)}^{p,n}|^2$ given by Eq. (\ref{line_eq2}) for some periodic orbits plotted in Figs. \ref{F1_fig:tray_clas} and \ref{fig4:tray_quantum}. 
The classical velocity components for $v(l)$ were obtained with the corresponding classical orbit. The positions of the segment lines defined by Eq. (\ref{line_eq2}) are dependent on the parameter $q_{r_{0},m_{0}}$ and the central state $(r_{0},m_{0})$. The quantum-classical connection implies that the value of $\gamma_{r_{0},m_{0}}$ must be very similar to the classical parameter $\gamma_{p,n}$. The fulfillment of these conditions ensures the lines intersect the boundary at the correct bounce points.

The value of $S$ has to be large enough to approximate the Dirichlet kernel to unity; this implies that $\Delta$ is as small as possible. 
Since $\Delta$ depends on the index $p$, for the orbit like $\mathcal{R}(8,3)$ setting $S=2$ was acceptable. On the other side, a higher order $r_{0}$ has the effect of sharpening the trajectories because, at the higher the order, the more similar the values of the classical parameter $\gamma_{p,n}$ and the eigenvalue $\gamma_{r_{0},m_{0}}$ of the central traveling state are. This is why we used a higher order ($r_0=700$) for the trajectory $\mathcal{L}(10,3)$ to improve its sharpness and visibility. Evaluating the zeros of the RMFs for such high values of $r_0$ becomes a computational task as it was necessary to calculate up to 100 zeros. We finalize by mentioning that the integral equation (\ref{line_eq2}) and the normalization constant were calculated numerically.

\section{Conclusions}

This work shows that a suitable superposition [\ref{coherentstate}] of nearly degenerate traveling states $\psi_{r,m}^{+}(\xi,\eta)$ of the elliptic billiard can be localized in classical periodic trajectories of the rotational and librational type. The superposition requires the precise evaluation of the parametric zeros of the radial Mathieu functions for very high orders and values of the parameter $q$. For this, we developed routines for calculating the RMFs based on different series of products of modified Bessel functions \cite{McLachlan,NISTBOOK}. We got accuracies of the order of $10^{-10}$ in evaluating RMFs for orders above 700.
The current density vector field $\mathbf{J}(\xi,\eta;\phi_{0})$ emulates the velocity flow of a particle in the classical regime; thus, classical trajectories can be visualized following the probability density flow. The phase distribution $\arg \mathbf{J}$ of the POQS shows the appearance of vortices of unit charge distributed throughout the billiard surface. In the case of rotational orbits, a set of in-line vortices occurs along the interfocal line. Their number depends on the order $r_0$ of the central eigenstate of the superposition. The interfocal vortex chain does not appear in the case of librational trajectories.
By applying the plane wave expansion of the eigenstates Eqs. (\ref{PWEx}), it was possible to simplify the general superposition of the POQS Eq. (\ref{line_eq2}). The simplified expression clearly depicts the straight segments of the classical trajectories. A sinc function gives the variation perpendicular to each segment of the trajectory. Line segments with higher sharpness could be obtained by adjusting the expression parameters.



%

\end{document}